\documentclass[aps,prx,twocolumn,nofootinbib,longbibliography,nobibnotes,10pt]{revtex4-2}

\usepackage[utf8]{inputenc} 
\usepackage[T1]{fontenc}  
\usepackage{lmodern}
\usepackage{graphicx,xcolor,overpic,mathtools}
\usepackage{amsmath,amssymb}
\usepackage{hyperref}
\usepackage{physics,tensor,cancel}
\usepackage{enumerate}
\usepackage{tikz}

\usepackage{subcaption}

\hypersetup{
    colorlinks=true,
    linkcolor=blue,
    citecolor=blue,
    urlcolor=blue,
    pdftitle={Your Review Title},
    pdfauthor={Your Name}
}

\newcommand{\hide}[1]{}
\DeclareMathOperator{\diag}{diag}
\newcommand{\prlsec}[1]{\noindent {\bf \em #1. }}
\newcommand{\ric}{\mathcal{R}}

\let\physicsqty\qty 
\let\qty\relax      
\usepackage{siunitx}
\let\qty\physicsqty 

\DeclareSIUnit{\MeV}{MeV}
\DeclareSIUnit{\fm}{fm}

\begin{document}

\title{Quantization of Brane-Skyrmions via Physics-Informed Neural Networks}
\author{Jose A. R. Cembranos}
\affiliation{Departamento de F\'isica Te\'orica and IPARCOS, Facultad de Ciencias F\'isicas, Universidad Complutense de Madrid, 28040 Madrid, Spain}

\author{Alberto García Martín-Caro}
\email{alberto.garcia.martin-caro@uvigo.gal}
\affiliation{Instituto de Física, Computación e Ciencias Aeroespaciais (IFCAE), Universidade de Vigo. 32004 Ourense, Spain}
\affiliation{
 Institute of Theoretical Physics, Jagiellonian University, Lojasiewicza 11, 30-348 Kraków, Poland
}

\author{Sergio S. Rentero}
\affiliation{Departamento de F\'isica Te\'orica and IPARCOS, Facultad de Ciencias F\'isicas, Universidad Complutense de Madrid, 28040 Madrid, Spain}

\begin{abstract}
In this work, we investigate the canonical quantization of topological solitons appearing in braneworld scenarios. In particular, we focus on Brane-Skyrmions, topological field configurations analogous to standard Skyrmions, which emerge as solutions of the 
Dirac–Nambu–Goto action supplemented by an induced curvature term. By quantizing the (iso)spin collective coordinates of the Brane-Skyrmion, we obtain a Hamiltonian that we solve perturbatively via an expansion in powers of $J^2$, in contrast to the standard Skyrme model.
Furthermore, we implement a Physics-Informed Neural Network (PINN) to determine the soliton profile that minimizes the energy, consistently incorporating the backreaction from the quantized spin degrees of freedom. We conclude with a discussion of the potential applications of this framework to the description of hadronic spectra. Our results highlight both the theoretical potential of brane-defect models and the growing role of neural network methods in theoretical physics.
\end{abstract}

\maketitle

\section{Introduction}
    The concept of extra spatial dimensions dates back to the pioneering works of Kaluza and Klein \cite{KALUZA_2018,Klein:1926tv}, who attempted to unify electromagnetism and gravity. In modern theoretical physics, this idea has evolved into the highly studied \emph{braneworld} scenario. In this framework, our observable universe is modeled as a four-dimensional hypersurface, the ``brane'', embedded within a higher-dimensional bulk space 
    \cite{Arkani-Hamed:1998jmv,Rubakov:1983bb}.

    The presence of a rigid brane in the bulk spontaneously breaks the translational invariance of the extra dimensions. According to Goldstone's theorem, this symmetry breaking gives rise to massless scalar fields on the brane, or ``branons'' \cite{Dobado:2000gr}. Physically, these branons represent the transverse fluctuations of the brane into the extra dimensions and own a rich phenomenology \cite{Alcaraz:2002iu, Cembranos:2003mr, Cembranos:2003fu, Cembranos:2004jp, Cembranos:2005jc, Cembranos:2005sr}. However, the mathematical structure of the target manifold in which these fields reside allows for more than just perturbative oscillations, it can also support stable configurations known as topological defects. Intuitively, these defects occur when the field configuration ``wraps'' around the internal target space (co-dimensions) in a way that cannot be continuously deformed back to the trivial vacuum.

    In this work, we are particularly interested in a specific type of topological soliton related to the Skyrme model \cite{Skyrme:1961vq,Perring:1962vs}. Originally formulated in the context of nuclear physics, the Skyrme model successfully describes baryons (such as protons and neutrons) as topological solitons emerging from the low energy dynamics of pion fields, where the baryon number is identified with the topological charge of the soliton \cite{Adkins:1983ya,Witten:1983tx}. Inspired by this success, we consider the framework developed in \cite{Dobado:2000gr}, which demonstrates that the low energy effective action for branons, derived from a Dirac-Nambu-Goto-type action with induced curvature terms, shares a profound mathematical similarity with the chiral effective Lagrangian of low energy QCD. As a consequence, this brane action also supports Skyrmion-like solutions, termed ``Brane-Skyrmions'' \cite{Cembranos:2001rp, Cembranos:2008kg}. From this point of view, our motivation for this work is to investigate wether we can also describe baryons as Brane-Skyrmions and how the predictions from the brane model differ from the standard Skyrme model.
    
    For that purpose, one must perform the quantization of these solitons, as there has only been a classical description of (approximate) solutions in the brane model \cite{Cembranos:2001rp}. As for the standard Skyrme model, quantization is achieved by introducing time dependent collective coordinates \cite{Adkins:1983ya}, allowing the soliton to rotate as a rigid body and, because of this, acquiring quantized spin and isospin states.

    We emphasize that, although our framework involves branes and extra dimensions, it should not be confused with holographic approaches to baryons where these can be interpreted as either solitons or wrapped brane states \cite{Witten:1998xy,Sakai:2004cn,Bolognesi:2013nja}, nor with higher dimensional generalizations of the Skyrme model \cite{Blanco-Pillado:2008coy,Gudnason:2014uha}. Instead, our construction is rooted in the effective field theory describing brane fluctuations, where the Skyrmionic sector emerges naturally from the low energy dynamics of branons. In this sense, our construction shares more similarity with earlier proposals such as \cite{Matsuda:2003hd,CarrilloGonzalez:2016lor}.

    We begin by providing an in-depth description of the Brane-Skyrmion model at the classical level in Section \ref{Sec.II}, where we introduce the standard notation and the underlying mathematical framework ---such as the symmetries of the model--- and other relevant results, including the induced metric for the brane and the mass functional. Furthermore, we implement Physics-Informed Neural Networks (PINNs) as the computational tool to overcome the highly non-linear differential equations governing the soliton profile. We end this section with a discussion on the physical properties of the classical Brane-Skyrmions through Derrick's scaling argument \cite{Derrick:1964ww}. Subsequently, we proceed to the canonical quantization of these solutions in Section \ref{sec:III} by introducing a rigid rotation ansatz for the soliton into the highly non-linear effective action of the brane. This allows us to construct the quantum Hamiltonian for Brane-Skyrmions and develop a systematic perturbative approach in a slow-rotation regime, ultimately revealing how the quantum centrifugal barrier stabilizes the soliton against collapse to a point-like defect. Finally, as a proof of concept, we perform a phenomenological fit of the quantized model to certain empirical hadronic observables.

    In this work we use the metric signature $(+,-,-,-)$, and for all but the last section, we employ natural units, in which $c=G=\hbar=1$.
    
\section{Classical configurations of static Brane-Skyrmions}
\label{Sec.II}
\prlsec{Effective action and mass}
    To maintain clarity in the subsequent mathematical framework, we introduce the following standard notation for the coordinates of the manifolds under consideration:
    \begin{itemize}
        \item $D$-dimensional bulk spacetime: $X^A$, with $A=0,\dots,D-1$.
        \item $d$-dimensional world-volume: $x^\mu$, with $\mu=0,\dots,d-1$. The component $x^0$ is typically identified as the coordinate time.
        \item $\bar{d}$-dimensional (with $\bar{d}=D-d$) target space: $X^I$, with $I=d,\dots,D-1$.
    \end{itemize}
    The effective action for the brane must inherit the bulk spacetime isometries through a set of global symmetries, some of wich may be nonlinearly realized (see e.g. \cite{deRham:2010eu,Garoffolo:2025igz}).
    

    On the other hand, gauge symmetry dictates that the effective action must also be invariant under reparametrizations of the worldvolume coordinates by arbitrary functions $\xi^\mu(x)$:
    \begin{equation}
        \delta_g X^A=\xi^\mu(x)\partial_\mu X^A.
    \end{equation}
    We can exploit this gauge freedom to select the so-called unitary gauge for parameterizing the brane's world-volume:
    \begin{equation}
        X^\mu(x)=x^\mu,\quad X^I(x)=\pi^I(x);\quad X^A(x)=(x^\mu,\pi^I(x)).
    \end{equation}
    Here, $\pi^I$ denotes the $\bar{d}$ fields representing the position of the brane along the extra dimensions. Physically, this implies that transverse fluctuations of the brane within the bulk space are perceived by an observer on the brane as a set of scalar fields. These correspond precisely to the  Nambu-Goldstone modes (or \emph{branons} \cite{Cembranos:2001rp}) associated with the spontaneous symmetry breaking of the original isometry group. 

The action governing the dynamics of the brane at lower energies will just include such fields, and must be invariant both under worldvolume reparametrizations and bulk isometry transformations. Reparametrization invariance implies that this action must be an arbitrary function of diffeomorphism-scalar quantities constructed out of the induced worldvolume metric,
\begin{equation}
    g_{\mu\nu}=\pdv{X^A}{x^\mu}\pdv{X^B}{x^\mu}\tilde{g}_{AB}=\tilde{g}_{\mu\nu}-\partial_\mu \pi^I\partial_\nu\pi_I.
    \label{eq:induced_g}
\end{equation}
where $\tilde{g}_{AB}$ is the background bulk metric.

The construction of this action can be guided by two different perspectives. From an effective field theory viewpoint, the low energy physics can be described by including an arbitrary number of terms ordered by increasing derivative power. Alternatively, if we require a framework valid at all energy scales, the form of the action must be further constrained by imposing that it only contains terms leading to second-order equations for the branons. Following this second approach, for single codimension branes, the terms appearing in the effective action were first described in \cite{deRham:2010eu}, called Dirac-Born-Infield (DBI) Galileons. The generalization of the probe brane construction for arbitrary codimension was developed in \cite{Hinterbichler:2010xn,Garoffolo:2025igz}.  It turns out that the \emph{unique} brane effective action in four dimensions for codimension $\bar d\geq 4$ satisfying such constraint is \cite{Trodden:2011xh}:
    \begin{equation}
        S=\int_{M_4}\dd[4]{x}\sqrt{g}(-f^4+\lambda f^2 \ric),
    \end{equation}
where $\mathcal{R}$ is the Ricci scalar of the induced metric, $\lambda$ is a coupling parameter, and $f$ represents the brane tension. In principle, their value can be obtained from the underlying ultraviolet theory that generates the brane \cite{Blanco-Pillado:2024bev,Aurrekoetxea:2026xrl}, but in the bottom-up approach to the effective action, they may be considered as free parameters.
The allowed four-dimensional
terms are then constrained to just two, namely, the cosmological constant term and the induced Einstein-Hilbert term. Rewriting this action in terms of the branon fields leads to a relativistically
invariant, multi-field generalization of DBI with second order field equations.

To construct a Skyrmion-like topological soliton within a physical 4-dimensional spacetime, the target space of the scalar fields must necessarily be 3-dimensional. Therefore, we consider a 3-brane with a four-dimensional worldvolume $M_4$ embedded within a bulk space. In principle, the total number of extra dimensions could be higher, provided that the low energy physics and the dominant phenomenological effects are effectively driven by exactly $\bar{d}=3$ dimensions. For this purpose, we choose these dimensions to be $S^3$, yielding a $D=7$ bulk $M_4\times S^3$ \cite{Cembranos:2001rp}. In this scenario, the original isometry group is then $\mathrm{ISO}(1,3) \times \mathrm{SO}(4)$. The presence of the brane on the compact dimensions breaks the isometry down to $\mathrm{ISO}(1,3) \times \mathrm{SO}(3)$, and hence the branon fields $\pi^I$ are mappings of the form $\pi^I: \mathbb{R}^3\to K$, with 
\begin{equation}
     K= \frac{\mathrm{ISO}(1,3) \times \mathrm{SO}(4)}{\mathrm{ISO}(1,3) \times \mathrm{SO}(3)} \cong \frac{\mathrm{SO}( 4)}{\mathrm{SO}(3)} \cong S^{3}.
\end{equation}
    Assuming the branon fields $\pi^I$ vanish at spatial infinity ($\pi^I\to 0$ as $r\to\infty$), the physical space $\mathbb{R}^3$ is effectively one-point compactified into $S^3$. Consequently, the fields constitute mappings $\pi^I:S^3\to S^3$, allowing them to be classified according to the third homotopy group of $K$, namely $\pi_3(K)=\pi_3(S^3)=\mathbb{Z}$. Thus, each mapping is characterized by an integer, which serves as a topological invariant. Intuitively, this number counts how many times the physical space ``wraps'' around the internal target space. Here, this target space is physically realized by the extra dimensions, which from the perspective of the hadronic model, effectively behave as the internal isospin space of the pions. Henceforth, we shall refer to this extra space simply as the internal space.
    The topological nature of the $\pi^I$ fields ensures that Brane-Skyrmions represent stable, finite energy configurations that cannot be continuously deformed into the trivial vacuum state ($\pi^I=0$).

    For static configurations, the Brane-Skyrmion mass is defined as:
    \begin{equation}\label{eq:mass}
        M_S[\pi]=\int_{M_3}\dd[3]{x}\sqrt{g}(-f^4+\lambda f^2 \ric)-M_S[0].
    \end{equation}
    This represents the energy strictly arising from the presence of the topological defect (the Brane-Skyrmion) on the brane, subtracting the infinite contribution from the flat brane.
\\
\\
\prlsec{Hedgehog ansatz} To simplify the calculations, spherical coordinates are introduced in both the $M_4$ and $K\sim S^3$ spaces. In $M_4$, we denote the coordinates as $\{t,r,\theta,\phi\}$, with $\phi\in[0,2\pi)$, $\theta\in[0,\pi)$, and $r\in[0,\infty)$. In the coset manifold $K$, the Cartesian coordinates are the branon fields, denoted by $\{\pi^I\}$. The spherical coordinates of $S^3$ are denoted as $\{\chi_K,\theta_K,\phi_K\}$, with $\phi_K\in[0,2\pi)$ and $\theta_K,\chi_K\in[0,\pi)$.

    Throughout this work, we will consistently assume the \emph{hedgehog ansatz}. This is a highly symmetric configuration that correlates the spatial directions with the internal field directions, causing the field to point radially outward from the center. Mathematically, the Brane-Skyrmion with winding number $n_W$ is given by the non-trivial mapping $\pi^I: S^3\to S^3$, defined by $\phi_K=\phi$, $\theta_K=\theta$, and $\chi_K=F(r)$, subject to the boundary conditions $F(0)=n_W\pi$ and $F(\infty)=0$. Therefore, the spherical coordinates in $S^3$ are related to the physical branon fields via \cite{Cembranos:2001rp}:
    \begin{align}\label{eq:hedgehog}
        &\pi^1(x)=v\sin F(r)\sin\theta\cos\phi, \nonumber\\
        &\pi^2(x)=v\sin F(r)\sin\theta\sin\phi, \\
        &\pi^3(x)=v\sin F(r)\cos\theta, \nonumber
    \end{align}
    where $v$ represents the radius of the 3-sphere.
    
    Using these coordinates, the induced metric on the brane takes the diagonal form
    \begin{equation}\label{eq:static-metric}
        g_{\mu\nu}=\diag\qty(1,-B(r),-\rho^2(r),-\rho^2(r)\sin^2\theta),
    \end{equation}
    where
    \begin{equation}\label{eq:B&rho}
        B(r)=1+\frac{v^2}{f^4}F'^2(r),\quad \rho^2(r)=r^2+\frac{v^2}{f^4}\sin^2 F(r).
    \end{equation}
    Throughout this paper, a prime is used to denote the derivative of a function with respect to its single argument.
    This metric is derived from the geometry of the bulk space, following \eqref{eq:induced_g}:
    \begin{equation}
        g_{\mu\nu}=\eta_{\mu\nu}-\frac{1}{f^4}h_{IJ}(\pi)\partial_\mu\pi^I\partial_\nu\pi^J,
        \label{eq:induced_metric_h}
    \end{equation}
   with $h_{IJ}(\pi)$ the metric of the coset space $K$ (a 3-sphere in our case), with signature $(+,+,+)$.
    
    Finally, the Brane-Skyrmion mass (\ref{eq:mass}) can be expressed as a functional of $F(r)$ simply by computing $\sqrt{g}$ and $\ric$ from the metric \cite{Cembranos:2001rp}:
    \begin{equation}\label{eq:static-det}
        \sqrt{g}=\sqrt{B(r)}\rho^2(r)\sin\theta,\quad \ric=-\frac{2}{\rho^2}\left(1-\frac{1}{A}\right)-\frac{2A'}{A^2\rho\rho'},
    \end{equation}
    where $A(r)=B(r)/\rho'^2(r)$.
    \\
    \\
\prlsec{Atiyah-Manton ansatz}
    Once the classical mass of the Brane-Skyrmion has been expressed as a functional of $F(r)$, $M_S[F]$, we can minimize it by means of conventional methods to obtain both the soliton profile $F(r)$ and the energy of the configuration. Before addressing the general case\footnote{Throughout this work, we use ``the general case'' to refer to $F(r)$ without specifying a particular ansatz. However, we keep working under the hedgehog ansatz and all previously established assumptions.}, we recall the Atiyah-Manton ansatz \cite{Atiyah:1989dq} for a $n_W=1$ skyrmion:
    \begin{equation}\label{eq:AM-ansatz}
        F(r)=\pi\left(1-\frac{1}{\sqrt{1+(L/r)^2}}\right),
    \end{equation}
    where $L$ is a parameter characterizing the size of the soliton.

    This ansatz, originally derived from the holonomy of Yang-Mills instantons in $\mathbb{R}^4$ \cite{Atiyah:1989dq}, accurately reproduces the solution of the original $SU(2)$ Skyrme model \cite{Skyrme:1961vq,Perring:1962vs,Adkins:1983ya}. Consequently, it was employed in \cite{Cembranos:2001rp} to approximate the Brane-Skyrmion profile, given their close resemblance to standard Skyrmions, as discussed in the introduction.

    By adopting the Atiyah-Manton ansatz (\ref{eq:AM-ansatz}), the mass of the soliton reduces to an ordinary function of $L$. Thus, the energy of the configuration can be determined simply by minimizing $M_S(L)$ with respect to $L$. The optimal value, denoted as $L_m$, depends on the parameter $\lambda$ (see Figure 4 in \cite{Cembranos:2001rp}):
    \begin{equation}
        \left.\dv{M_S}{L}\right|_{L_m}=0 \;\to\; M_S\equiv M_S(L_m),\; L_m=L_m(\lambda).
    \end{equation}
    While this ansatz has been studied in the context of Brane-Skyrmions in \cite{Cembranos:2001rp}, it has not yet been rigorously proven that the solutions obtained by minimizing the Atiyah-Manton ansatz serve as a sufficiently accurate approximation to the real minimization. In fact, we will now show that this ansatz is not very accurate in  the $\lambda\to 0$ limit.
    \\
    \\
\prlsec{Numerical solution}
    As noted in \cite{Cembranos:2001rp}, minimizing the general case without assuming the Atiyah-Manton ansatz for the profile function (\ref{eq:AM-ansatz}) involves solving a complicated second order differential equation. Although it can be solved with standard numerical methods given sufficient computational effort, we opt to introduce a more powerful approach that will remain advantageous when addressing the quantum case: Physics Informed Neural Networks (PINNs).

    Unlike standard machine learning models that require large datasets to learn interpolations (further details are given in Appendix \ref{appendix:B}), PINNs can act as unsupervised differential equation solvers. By evaluating the gradients of the neural network with respect to the input coordinates via automatic differentiation, the network optimizes its weights to strictly minimize the action functional.

    To implement a PINN for our minimization problem, we begin by defining the architecture of the network using the following expression for the soliton profile, $F(r)$:
    \begin{equation}\label{eq:nn-ansatz}
        F(r)=F_0(r)+V(r)N(r),
    \end{equation}
    where:
    \begin{itemize}
        \item $F_0(r)$ is an initial ansatz for the solution that satisfies the boundary conditions $F_0(0)=\pi$ and $F_0(\infty)=0$. The Atiyah-Manton ansatz (\ref{eq:AM-ansatz}), for instance, can serve this purpose.
        \item $N(r)$ is the output of the neural network, parameterized by its weights and biases. This function modifies the initial ansatz, $F_0(r)$, in order to reach the exact solution $F(r)$ that minimizes the energy functional.
        \item $V(r)$ is a weight function for the variation of $F_0(r)$, satisfying the boundary conditions $V(0)=V(\infty)=0$. The following ansatz seems to give reasonable results for our purposes:
        \begin{equation}
            V(r)=\frac{L_m}{L_m+r}e^{-\epsilon^2/r^2}e^{-\epsilon^2/(r-r_\text{max})^2},
        \end{equation}
        where $r_\text{max}$ is the numerical integration cutoff, $\epsilon$ is a length scale that determines the steepness of the bump function, and $L_\text{m}$ is the optimal length scale found via minimization of the energy using the Atiyah-Manton ansatz; it determines the characteristic size of the soliton. For this work, we have used $\epsilon=10^{-5}$ since it is the value that better reproduces traditional methods.
    \end{itemize}

    Having established the architecture, we outline the workflow of this particular model, which closely resembles a gradient flow method. The input for the NN consists of a set of values for the coordinate $r$. It is fed a column vector with $N_r$ points distributed between $0$ and the maximum value of the radial coordinate, $r_{\rm max}$ that we wish to evaluate (extending to infinity is computationally prohibitive). We define two hidden layers comprising 40 neurons each. On each layer, a Gaussian Error Linear Unit (GELU) activation function is applied, providing a smooth non‑linear transformation\footnote{The smoothness and differentiability of the activation function are important to ensure that derivatives of the network output via automatic differentiation are well defined and numerically stable. See \cite{Wang_2023} for a systematic analysis of the effects of activation‑function choices in PINNs.}.  Each hidden layer contains weights and biases (the network parameters, $\theta_i$) that are updated after every iteration.
    The output of the neural network is the function $N(r)$ introduced in (\ref{eq:nn-ansatz}). The mass functional (\ref{eq:mass}) is evaluated using the function $F(r)$ constructed from the neural network output. This yields the energy of the configuration. On each iteration, the model calculates the gradient of the resulting energy with respect to each parameter $\theta_i$ of the neural network, $\pdv*{M_S}{\theta_i}$. The optimizer dictates the direction in which the network parameters should be updated in the subsequent iteration to minimize $M_S$. For our purposes, the ``Adam'' optimizer is sufficient.

    The implementation of this PINN, utilizing the \texttt{PyTorch} library in Python, has been made publicly available at \cite{Github_code}. The original code is loosely based on some lecture notes on PINNs for strong gravity and PDE solving by R. Luna \cite{R_Luna_lectures}.

    Let us now examine the numerical results yielded by the PINN. Note that the convergence to a global minimum of the energy is not ensured by this method. To check the reliability of these results, without loss of continuity, a validation for the static case is provided in Appendix \ref{appendix:A}. In Figures \ref{fig:NN_1} and \ref{fig:NN_001}, one can observe the solutions obtained for two different values of the rescaled parameter $\lambda^*=\lambda/f^2 R_B^2$. The profiles obtained by the PINN are also compared to those obtained via a shooting method.
    
    Note that the solution becomes point-like as $\lambda\to 0$ (see the next subsection); yet, it remains topologically non-trivial, as its mass satisfies a topological bound 
    \cite{Cembranos:2001rp}
    \begin{equation}
        M_S\geq V(S^3)f^4 |n_W|
    \end{equation}
which is saturated precisely in the $\lambda=0$ case. In fact, the curvature term contributes to an upper bound on the mass for positive $\lambda$:
\begin{equation}
  2\pi^2R^3f^4 \leq M_S\leq 2\pi^2R^3f^4\qty(1+\lambda\frac{6}{f^2 R^2}),
\end{equation}
since the largest Ricci curvature is achieved in the case of pointlike skyrmions, namely, when all the wrapping of the brane takes place in a single point. These configurations correspond to the so-called ``wrapped states'' in \cite{Cembranos:2001rp}. 

\begin{figure}[ht!]
    \centering

    \begin{subfigure}{\linewidth}
        \centering
        \includegraphics[width=\linewidth]{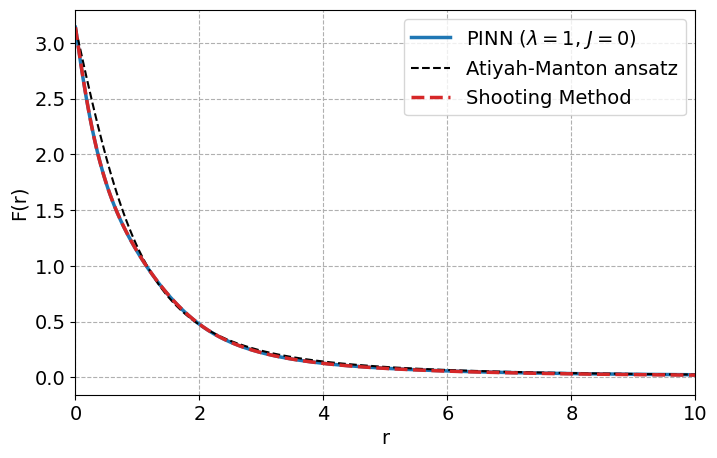}
        \caption{$\lambda^* = 1$}
        \label{fig:NN_1}
    \end{subfigure}

    \vspace{0.5em}

    \begin{subfigure}{\linewidth}
        \centering
        \includegraphics[width=\linewidth]{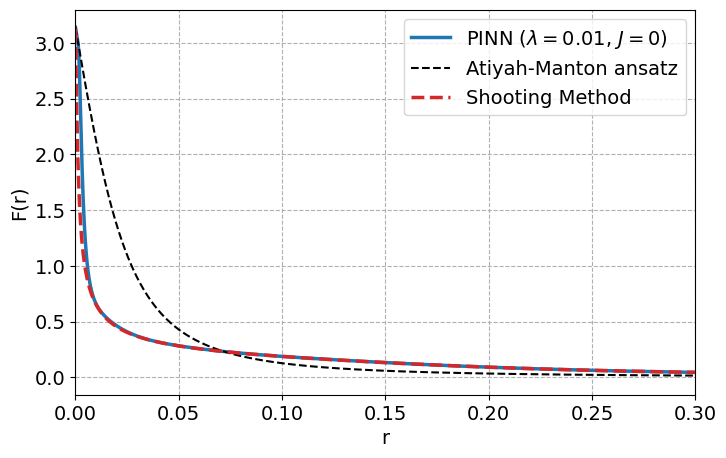}
        \caption{$\lambda^* = 0.01$}
        \label{fig:NN_001}
    \end{subfigure}

    \caption{Neural network solutions for a static ($J=0$) Brane-Skyrmion for two values of $\lambda^*$, compared to the Atiyah-Manton and Shooting method solutions.}
    \label{fig:NN_combined}
\end{figure}

    The resulting solution for the profile function $F(r)$ is shown for $\lambda=\{0.01,1\}$, respectively, in Figs. \ref{fig:NN_1} and \ref{fig:NN_001}. We see that, although for values on the order of $\lambda^*\sim 1$ the Atiyah-Manton ansatz (\ref{eq:AM-ansatz}) provides a good approximation to the exact solution for static Brane-Skyrmions (see Figure \ref{fig:NN_1}), they differ for small values of $\lambda^*$ (see Figure \ref{fig:NN_001}),  when the soliton tends to become point-like. 

    It is interesting to analyze the situation by using scaling arguments \cite{Derrick:1964ww, Manton:2004tk}. The total energy, $E(L)$, of a soliton changes under a rescaling of its profile: $F(r)\to F(r/L)$. In the Skyrme model, the total energy scales as \cite{Manton:2004tk}:
    \begin{equation}
        E_\text{Skyrmion}(L)=E_2 L+E_4L^{-1},
    \end{equation}
    where $E_2$ and $E_4$ are constants. 
    
    As $L\to 0$, the $E_4$ term diverges, making it impossible for a stable Skyrmion to collapse to a point. Conversely, as $L\to\infty$, the energy also diverges; thus, a stable configuration must possess a non-zero, finite size, a desirable feature for the description of nucleons. The Atiyah-Manton ansatz (\ref{eq:AM-ansatz}) is specifically tailored to describe such a configuration. 
    
    However, this argument yields different implications for Brane-Skyrmions. Let us take the mass functional (\ref{eq:mass}) in the limit $\lambda\to 0$ (i.e. neglecting the contribution from the induced curvature of the brane) and apply the scaling $r=L\tilde{r}$:
    \begin{equation}
        E_\text{BS}(L)\propto f^4\int\dd{\tilde{r}} L\left[\sqrt{1+\frac{R_B^2}{L^2}F'^2}(L^2\tilde{r}^2+R_B^2\sin^2 F)\right],
        \label{eq:E_dens}
    \end{equation}
    where $R_B=v/f^2$ represents the characteristic scale for the size of the soliton, and $F'\equiv\dv*{F}{\tilde{r}}$, which introduces the additional $1/L$ factor.

    In the limit $L\to 0$, the square root can be approximated simply as $\tfrac{R_B}{L}|F'|$, which cancels the overall $L$ factor and yields a finite energy:
    \begin{equation}
        \lim_{L\to 0} E_\text{BS}(L)\propto \int_0^\infty \dd{\tilde{r}} f^4 R_B^3 F'\sin^2 F >0.
    \end{equation}
    Consequently, it is possible for a Brane-Skyrmion to become point-like while maintaining a finite energy. 
    Finally, this analysis provides a simple derivation of the topological bound for the mass found in \cite{Cembranos:2001rp} for the first time:
    \begin{equation}
        M_S(\lambda\to 0)=4\pi f^4 R_B^3\int_0^\pi\dd{u}\sin^2 u=2\pi^2 f^4 R_B^3,
    \end{equation}
    where the change of variable $u=F(\tilde{r})$ allows to perform the radial integral without requiring the explicit form of the profile.

\section{Quantization of spherical Brane-Skyrmions}
\label{sec:III}
    Up to this point, we have reviewed several known properties of Brane-Skyrmions, which were previously discussed in \cite{Cembranos:2001rp}, alongside analyzing new physical implications. However, these are strictly static configurations, lacking any notion of angular momentum. Furthermore, we have demonstrated that these static solitons admit point-like configurations. While such characteristics could potentially model spinless, scalar point particles, they fail to describe baryons or particles with internal structure and spin. Consequently, the quantization of Brane-Skyrmions becomes necessary.

    To simplify the inherently complex calculations, we will maintain the hedgehog ansatz (\ref{eq:hedgehog}), restricting our quantization to spherical configurations of Brane-Skyrmions (i.e., those that can be fully described by the one dimensional profile $F(r)$).
    Further, we take the same approach as for the canonical quantization of standard Skyrmions \cite{Adkins:1983ya} and quantize the zero (iso-)rotational modes.
    Indeed, the induced metric \eqref{eq:induced_metric_h} is manifestly invariant under global $SO(3)$ rotations, $\pi^I=R^I_J\pi^J$, where $R\in SO(3)$ is independent of the space-time coordinates. Consequently, the effective action for Brane-Skyrmions is globally $SO(3)$-invariant, meaning that rotating a static solution incurs no potential energy cost. These continuous symmetries correspond to zero modes of the classical system, which can be promoted to dynamical collective coordinates by making the transformation parameters time-dependent:
    \begin{equation}\label{eq:rigid-ansatz}
        \pi^I(t)=R^I_J(t)\pi_0^J,\quad R(t):=e^{t\omega_i L^i},
    \end{equation}
    where $\pi_0^I$ denotes the static branon fields\footnote{Note that both $\pi^I(t)$ and $\pi_0^I$ also depend on the spatial coordinates $\{r,\theta,\phi\}$. We will omit this explicit dependence for the rest of this work for simplicity.} (\ref{eq:hedgehog}), $L^i\in \mathfrak{so}(3)$ are the generators of the $SO(3)$ group, and $\omega_i$ are three real parameters.

    The expression adopted for $R(t)$ can be physically interpreted as a rigid body rotation ansatz, where $\omega_i$ represents the components of the angular velocity along the three spatial axes. Later, we will assume a slow-rotation regime, so that this rigid ansatz is justified.
\\
\\
\prlsec{Effective Lagrangian}
     Let us compute the new effective Lagrangian under the rotation ansatz (\ref{eq:rigid-ansatz}). While this could be performed using a general time-dependent rotation, we will maintain the rigid rotation ansatz for simplicity. Note that, since there are no second-order time derivatives in the action, we can simply substitute $\omega^i\equiv\dot{\theta}^i(t)$ at each step of the derivation, where $\theta^i(t)=t\omega^i$ describes the rigid rotation. 

    The first step is to compute the components of the induced metric on the brane. We can express it as:
    \begin{equation}
        g_{\mu\nu}=\eta_{\mu\nu}-G_{\mu\nu}(\pi),\quad G_{\mu\nu}(\pi)=\frac{1}{f^4}h_{IJ}(\pi)\partial_\mu\pi^I\partial_\nu\pi^J,
        \label{G_munudef}
    \end{equation}
    where $\pi^I\equiv\pi^I(t)$ now represents the dynamical field configurations.
    
    Let us explicitly compute the metric $h_{IJ}(\pi)$ in terms of the Cartesian coordinates $\{\pi^I\}$ of the coset space. Given that the coset space is isomorphic to $S^3$, we can embed it into $\mathbb{R}^4$ using coordinates $(Y^0,\pi^I)$ subject to the following constraint:
    \begin{equation}
        (Y^0)^2+\pi^2=v^2 \quad\Rightarrow\quad \dd{Y^0}=-\frac{\pi_I\dd{\pi^I}}{\sqrt{v^2-\pi^2}},
    \end{equation}
    where $v$ is the radius of the 3-sphere, and $\pi^2\equiv\pi_I\pi^I$.
    
    Consequently, the line element of the coset space is given by:
    \begin{equation}
        \dd{S_K}^2=\delta_{IJ}\dd{\pi^I}\dd{\pi^J}+\frac{\pi_I\dd{\pi^I}\pi_J\dd{\pi^J}}{v^2-\pi^2},
    \end{equation}
    yielding the metric tensor:
    \begin{equation}
        h_{IJ}(\pi)=\delta_{IJ}+\frac{\pi_I\pi_J}{v^2-\pi^2}.
    \end{equation}
    We now proceed to calculate the different components of $G_{\mu\nu}(\pi)$ as defined in \eqref{G_munudef}.
    On the one hand, the purely spatial components $G_{ij}(\pi)$ remain invariant under time-dependent rotations, as the time dependence exclusively affects components involving temporal derivatives.
    
        \hide{Expanding it explicitly yields:
        \begin{align}
            G_{ij}(\pi)
            &=\frac{1}{f^4}h_{IJ}(\pi)\partial_i\pi^I\partial_j\pi^J \nonumber\\
            &=\frac{1}{f^4}h_{IJ}(\pi)R^I_K R^J_L\partial_i\pi_0^K\partial_j\pi_0^L \nonumber\\
            &=\frac{1}{f^4}h_{KL}(\pi_0)\partial_i\pi_0^K\partial_j\pi_0^L=G_{ij}(\pi_0),
        \end{align}
        where we have utilized the following identities:
        \begin{equation}
            \delta_{IJ}R^I_K R^J_L=\delta_{KL},\quad \pi_I R^I_J=\pi_{0,J},\quad \pi^2=\pi_0^2.
        \end{equation}}
        
        Therefore, in the coordinate system $\{r,\theta,\phi\}$, the spatial metric assumes the same form as in the static case (\ref{eq:static-metric}):
        \begin{equation}
            g_{ij}=\diag\qty(-B(r),-\rho^2(r),-\rho^2(r)\sin^2\theta).
        \end{equation}
        On the other hand, to evaluate the $G_{00}(\pi)$ component, we must first compute $\partial_t R$. Using (\ref{eq:rigid-ansatz}):
        \begin{equation}
            \partial_t R^I_J=\omega^i(L_i)^I_K R^K_J=\tensor{\epsilon}{^I_i_K}\omega^i R^K_J,
        \end{equation}
        where we have adopted the following representation of the generators:
        \begin{equation}
            (L_i)^I_J=\tensor{\epsilon}{^I_i_J},\quad [L_i,L_j]=\epsilon_{ijk}L^k.
        \end{equation}
        Note that the indices $I,J,K,\dots$ refer to the internal coset space, whereas the indices $i,j,k,\dots$ refer to the spatial manifold $M_3$ (which corresponds to $\mathbb{R}^3$). This leads to the following expressions:
        \begin{align}
            &\dot{\pi}^I=\partial_t R^I_J\pi^J_0=\tensor{\epsilon}{^I_i_J}\omega^i\pi^J,\\
            &\pi_I\dot{\pi}^I=\tensor{\epsilon}{^I_i_J}\pi_I\omega^i\pi^J=0,\\
            &\dot{\pi}_I\dot{\pi}^I=\tensor{\epsilon}{_I_i_J}\tensor{\epsilon}{^I_j_K}\omega^i\pi^J\omega^j\pi^K=\omega^2\pi^2-(\omega^i \pi_i)^2,
        \end{align}
        where we have made use of the total antisymmetry of the Levi-Civita tensor $\tensor{\epsilon}{_i_j_k}$ alongside the identity:
        \begin{equation}
            \tensor{\epsilon}{_I_i_J}\tensor{\epsilon}{^I_j_K}=\delta_{ij}\delta_{JK}-\delta_{iK}\delta_{Jj}.
        \end{equation}
        Consequently, we obtain:
        \begin{equation}
            G_{00}(\pi)=\frac{1}{f^4}h_{IJ}(\pi)\dot{\pi}^I\dot{\pi}^J=\frac{1}{f^4}\dot{\pi}_I\dot{\pi}^I\equiv\frac{\dot{\pi}^2}{f^4}.
        \end{equation}

        Finally, the mixed components $G_{0i}(\pi)$ can be computed in an analogous way, yielding:
        \begin{equation}
            G_{0i}(\pi)=\frac{1}{f^4}h_{IJ}(\pi)\dot{\pi}^I\partial_i\pi^J=\frac{1}{f^4}\dot{\pi}_I\partial_i\pi^I.
        \end{equation}
        and thus: $g_{0i}=-\tfrac{1}{f^4}\dot{\pi}_I\partial_i\pi^I$.
    Gathering these results, the induced metric on the brane assumes the following form:
    \begin{equation}
        g_{\mu\nu}=\mqty(1-\tfrac{\dot{\pi}^2}{f^4} & -\tfrac{1}{f^4}\dot{\pi}_I\partial_i\pi^I \\ -\tfrac{1}{f^4}\dot{\pi}_I\partial_i\pi^I & g_{ij}(\pi_0)).
        \label{eq:gmunu_general}
    \end{equation}
   We remark that \eqref{eq:gmunu_general} remains valid independently of the field ansatz. We will now impose both assumptions, (\ref{eq:hedgehog}) and (\ref{eq:rigid-ansatz}), to derive a more explicit version of this metric. Without loss of generality, we can align the axis of rotation with the $z$-axis. Indeed, since the hedgehog ansatz is spherically symmetric, the resulting inertia tensor is purely isotropic ($I_{ij}=\mathcal{I}\delta_{ij}$). Therefore, any arbitrary choice of rotation axis will yield an identical integrated Lagrangian. We set:
    \begin{equation}
        \boldsymbol{\omega}=(0,0,\omega),\quad R(t)=\mqty(\cos(\omega t) & -\sin(\omega t) & 0 \\ \sin(\omega t) & \cos(\omega t) & 0 \\ 0 & 0 & 1).
    \end{equation}
    
    Taking these conditions into account:
    \begin{align}
        &\dot{\pi}^2=\omega^2\pi^2-(\omega^i\pi_i)^2=v^2\omega^2\sin^2 F\sin^2\theta,\\
        &\dot{\pi}_ I\partial_i\pi^I=\epsilon_{IjJ}\omega^j\pi^J\partial_i\pi^I=v^2\sin^2 F\omega\sin^2\theta\delta_i^\phi,
    \end{align}
    where we have used the facts that $\pi_3=\pi_{0,3}$ and $\pi_1\partial_i\pi_2-\pi_2\partial_i\pi_1=\pi_{0,1}\partial_i\pi_{0,2}-\pi_{0,2}\partial_i\pi_{0,1}$. Note that the hedgehog ansatz (\ref{eq:hedgehog}) applies exclusively to the static fields $\pi_0$.
    
    This yields the explicit induced metric:
    \begin{equation}
        g_{\mu\nu}=\mqty(1-\omega^2\Lambda(r)\sin^2\theta & 0 & 0 & -\omega\Lambda(r)\sin^2\theta \\ 0 & -B(r) & 0 & 0 \\ 0 & 0 & -\rho^2 & 0 \\ -\omega\Lambda(r)\sin^2\theta & 0 & 0 & -\rho^2\sin^2\theta),
    \end{equation}
    where
    \begin{equation}
        \Lambda(r)=\rho^2(r)-r^2=\frac{v^2}{f^4}\sin^2 F(r).
    \end{equation}
    The non-vanishing $g_{0\phi}$ component reflects a frame-dragging effect in the azimuthal direction, implying a dynamical coupling between the soliton's internal degrees of freedom and its rotational motion.
    
    Our goal is now to compute the invariant volume element $\sqrt{g}$ and the scalar curvature $\ric$. The metric determinant is straightforward to calculate:
    \begin{equation}
        g=g_0\left(1-\frac{r^2\Lambda(r)}{r^2+\Lambda(r)}\omega^2\sin^2\theta\right),
    \end{equation}
    where $g_0=g_{rr}g_{\theta\theta}g_{\phi\phi}$ represents the determinant for the static Brane-Skyrmion metric.

    Consequently, we can write:
    \begin{equation}\label{eq:dynamic-metric}
        \sqrt{g}=\sqrt{B(r)}\rho^2(r)\sin\theta\sqrt{1-C(r)\omega^2\sin^2\theta},
    \end{equation}
    where $B(r)$ and $\rho^2(r)$ are given by (\ref{eq:B&rho}), and
    \begin{equation}\label{eq:C(r)}
        C(r)=\frac{r^2\Lambda(r)}{r^2+\Lambda(r)}=\frac{v^2}{f^4}\frac{r^2\sin^2 F(r)}{r^2+\tfrac{v^2}{f^4}\sin^2 F(r)}.
    \end{equation}
    Thus, $\sqrt{g}$ depends explicitly on $\omega^2$, a result perfectly consistent with the spherical symmetry inherently imposed by the hedgehog ansatz.
    
    Conversely, analytically computing the scalar curvature $\ric$ by hand is prohibitively complex. Therefore, we are led to use symbolic computation with Wolfram \texttt{Mathematica}. The resulting expression is extremely lengthy and is thus omitted here, but the interested reader can access the notebook with the result at \cite{Github_code}.

    Finally, the effective Lagrangian density for dynamical Brane-Skyrmions formally retains the structure of the static case:
    \begin{equation}
        \mathcal{L}_\text{eff}=(-f^4+\lambda f^2 \ric)\sqrt{g},
    \end{equation}
    where $\sqrt{g}$ is now given by (\ref{eq:dynamic-metric}) and $\ric$ by \texttt{Mathematica}'s result.
\\
\\
\prlsec{Canonical quantization}
    Once we have derived the effective Lagrangian in terms of $\omega^2$, which represents the dynamical degree of freedom we intend to quantize using the collective coordinates formalism \cite{Adkins:1983ya}, we proceed by computing the canonical momentum conjugate to $\omega$, which corresponds to the angular momentum of the Brane-Skyrmion:
    \begin{equation}
        J=\pdv{L(\omega)}{\omega},
    \end{equation}
    where $L(\omega)$ is the total effective Lagrangian:
    \begin{equation}\label{eq:dynamic-lagrangian}
        L(\omega)=\int\dd[3]{x}\sqrt{g}(-f^4+\lambda f^2 \ric).
    \end{equation}
    Note that we must use the total integrated Lagrangian, not merely the Lagrangian density. Using the density would not yield the correct canonical conjugate of $\omega$, leading to an incorrect Hamiltonian, which is the primary object of our quantization scheme.

    Consequently, if we can invert the relation $J=J(\omega)$ to obtain $\omega=\omega(J)$, it becomes straightforward to construct the Hamiltonian of the system in terms of the canonical momentum $J$:
    \begin{equation}
        H(J)=J\omega(J)-L[\omega(J)].
    \end{equation}
    However, recall that $L(\omega)$ depends on $\omega^2$ through a square root within an integrand that is evaluated over all space. It is analytically impossible to invert the relation $J=J(\omega)$ exactly. We are therefore led to rely on the previously assumed rigid rotation regime (\ref{eq:rigid-ansatz}), which dictates that $\omega$ must be a small quantity. Consequently, we can expand $L(\omega)$ around $\omega=0$ in powers of $\omega^2$:
    \begin{equation}
        L(\omega)=-\alpha+\beta\omega^2+\gamma\omega^4+\mathcal{O}(\omega^6),
    \end{equation}
    where $\alpha$, $\beta$, $\gamma$, and all higher-order coefficients correspond to distinct positive-definite integrals over space involving terms from $\sqrt{g}$ and $R$ (further details are given in Appendix \ref{appendix:C}, though it may be skipped without loss of continuity). Numerically, this series seems to converge in most cases of interest.
    
    The angular momentum is then given by:
    \begin{equation}
        J=\pdv{L(\omega)}{\omega}=2\beta\omega+4\gamma\omega^3+\mathcal{O}(\omega^5).
    \end{equation}
    In general, there is no exact method to invert this relation. However, since we are working under the assumption that $\omega$ is small, $J$ must also be small, allowing us to postulate an inverse series expansion: $\omega=\omega_1 J+\omega_3 J^3+\cdots$. Substituting this ansatz back into the previous equation and matching coefficients order by order, we find:
    \begin{equation}
        \omega=\frac{1}{2\beta}J-\frac{\gamma}{4\beta^4}J^3+\mathcal{O}(J^5).
    \end{equation}
    Finally, we perform the Legendre transformation to obtain the Hamiltonian:
    \begin{equation}\label{eq:classical-H}
        H(J)=\alpha+\frac{1}{4\beta}J^2-\frac{\gamma}{16\beta^4}J^4+\mathcal{O}(J^6).
    \end{equation}
    This procedure can be executed systematically, allowing us to compute the Hamiltonian to arbitrarily high orders, provided one can expand the Lagrangian to the corresponding order. An implementation of the method in \texttt{Mathematica} is available at \cite{Github_code}. We insist that \eqref{eq:classical-H} constitutes an approximate result in terms of a series whose convergence must be carefully studied (further details are given in Appendix \ref{appendix:B}). Furthermore, upon inspecting (\ref{eq:classical-H}) up to order $\mathcal{O}(J^2)$, it becomes apparent that $\beta$ is (up to a factor of 2) the leading order contribution to the moment of inertia of the soliton.
    
    Having formulated the Hamiltonian for classical Brane-Skyrmions as a power series in $J^2$, its quantization becomes straightforward. We simply promote $J^2$ to a quantum mechanical operator, $\hat{J}^2$, with well-established eigenvalues and eigenstates:
    \begin{equation}
        \hat{J}^2\ket{j}=j(j+1)\ket{j},\, \ket{j}\in\mathcal{H},
    \end{equation}
    where $\mathcal{H}$ is the Hilbert space spanned by the eigenstates of the operator $\hat{J}^2$, labeled by the quantum number representing the total angular momentum of the soliton, $j$.

    Note that $\ket{j}$ are simultaneously eigenstates of any power of $\hat{J}^2$:
    \begin{equation}
        (\hat{J}^2)^n\ket{j}=j(j+1)(\hat{J}^2)^{n-1}\ket{j}=\cdots=[j(j+1)]^n\ket{j}.
    \end{equation}
    Taking this into consideration, we promote the classical Hamiltonian to a corresponding quantum operator acting on $\mathcal{H}$:
    \begin{equation}
        \hat{H}=\sum_{n=0}^\infty H_{2n}(\hat{J}^2)^n,
    \end{equation}
    where the factors $H_{2n}$ denote the coefficients of the classical series expansion (further details are given in Appendix \ref{appendix:C}, though it may be skipped without loss of continuity).

    Finally, the energy of the soliton is evaluated as the eigenvalue of the quantum mechanical Hamiltonian:
    \begin{equation}\label{eq:quantum-H}
        \hat{H}\ket{j}=E_j\ket{j}=\sum_{n=0}^\infty H_{2n}(\hat{J}^2)^n\ket{j}=\sum_{n=0}^\infty H_{2n}[j(j+1)]^n\ket{j},
    \end{equation}
    yielding the final expression for the quantum mechanical energy eigenvalue:
    \begin{equation}\label{eq:energy}
        E_j[F]=\sum_{n=0}^\infty H_{2n}[F][j(j+1)]^n.
    \end{equation}
    We must bear in mind that the energy of the soliton remains a functional of its profile $F(r)$. Furthermore, the model contains two free parameters in addition to $R_B$ and $f$: the total angular momentum $j$ and the effective coupling $\lambda$. The value of $j$ dictates the specific particle state under description (e.g., $j=1/2$ for a proton/neutron, $j>1/2$ for higher resonances). Therefore, for each specified value of $j$, we obtain a distinct energy functional that must be minimized with respect to $F(r)$ for a given $\lambda$.

    Lastly, it is important to briefly address the convergence of the energy series (\ref{eq:energy}). Mathematically, since the Lagrangian series $L(\omega)$ converges rapidly in the slow-rotation regime, the analytic inverse function theorem guaranties that the classical Hamiltonian $H(J)$ also converges for sufficiently small $J$. However, upon canonical quantization, the continuous classical variable $J^2$ is replaced by discrete quantum eigenvalues $j(j+1)$. Consequently, this perturbative expansion remains physically rigorous only for low-lying spin states, as highly excited states (large $j$) would dynamically deform the soliton beyond the initial assumption of a spherically symmetric ansatz.
    \\
    \\

\prlsec{Numerical results and Experimental fit}

    To obtain numerical results, we must minimize the energy functional \eqref{eq:energy} for specific values of $\lambda$ and $j$. However, both parameters scale with the undetermined combination $fR_B$, so we need to introduce the following rescaled dimensionless parameters:
    \begin{equation}\label{eq:J-lambda}
        \lambda^*=\frac{\lambda}{f^2 R_B^2},\quad J^*=\frac{J}{f^4 R_B^4},
    \end{equation}
    where $J\equiv\sqrt{j(j+1)}$. The scaling powers $(fR_B)^n$ emerge naturally when one works with the dimensionless energy: $E^*=E/f^4 R_B^3$. Consequently, $J^*$ acts as an unconstrained parameter for the PINN, not as the physical spin of the nucleon. If we want to obtain numerical results related to physical observables, we should fix our free parameters $\{f,R_B,J^*,\lambda\}$ by fitting experimental data. In this case, we will use the nucleon mass $M_N$ and the nucleon isoscalar root-mean-square radius $r_N$, fixing the spin to $J_N=\sqrt{3/4}$ ($j_N=1/2$). This will allow use to fix three out of four parameters: $f$, $R_B$ and $J^*$. We will fix $\lambda^*$ previous to the fit, since leaving it as an undetermined parameter would make the problem much more complex.
    
    While the nucleon mass can be identified with the energy \eqref{eq:energy} of the soliton configuration, the theoretical isoscalar radius is defined via the baryonic density $\rho_B(r)$:
    \begin{equation}
        \langle r^2\rangle_{I=0}=\int_0^\infty\dd{r}r^2\rho_B(r).
    \end{equation}
    The baryonic density can be uniquely identified by requiring its volume integral to yield the total baryon number, which in our case must be $1$:
    \begin{equation}
        B=\int_0^\infty\dd{r}\rho_B(r)=1.
    \end{equation}
    In fact, as with the standard Skyrme model, the baryon charge can be identified to the topological winding number of the Brane-Skyrmion under consideration, setting $B\equiv n_W$\cite{Adkins:1983ya}. Therefore, the baryon number is determined exclusively by the mapping topology, and can be expressed as \cite{Manton:2004tk}:
    \begin{equation}
        B=\frac{1}{2\pi^2}\int_{\mathbb{R}^3}\dd{V}^*=-\frac{2}{\pi}\int_0^\infty\dd{r}F'(r)\sin^2 F(r),
    \end{equation}
    where $\dd{V}^*$ denotes the pullback of the volume form $\dd{V}$ from the target 3-sphere. The minus sign is introduced conventionally to ensure $B=1$ under the boundary conditions: $F(\infty)=0$ and $F(0)=\pi$.

    Therefore, the radial baryonic density can be readily identified:
    \begin{equation}
        \rho_B(r)=-\frac{2}{\pi}F'(r)\sin^2 F(r).
    \end{equation}
    Notice that this density coincides exactly with the one derived in the original Skyrme model \cite{Adkins:1983ya}, as it stems purely from the topological winding of the hedgehog mapping rather than the specific dynamical metric of the brane.

    To match the dimensionless PINN outputs to empirical data, we define the dimensionless energy $E^*=E/f^4 R_B^3$ and radius $r^*=\frac{1}{R_B}\sqrt{\langle r^2\rangle_{I=0}}$. We must impose the system of equations: $M_N=E(J_N,\lambda)$, $r_N^2=\langle r^2\rangle_{I=0}(J_N,\lambda)$, alongside the scaling relation $J_N=f^4 R_B^4J_N^*$. In principle, these four equations are sufficient to fix the three free parameters ($f$, $R_B$, and $J^*$, given a previously fixed $\lambda^*$). However, because this system is highly non-linear with respect to $J^*$, a direct evaluation is computationally prohibitive. A more robust algebraic strategy is required, so let us define the following combination:
    \begin{equation}
        \frac{M_N r_N}{J_N}=\frac{E^*(J_N^*,\lambda^*)r^*(J_N^*,\lambda^*)}{J_N^*}.
    \end{equation}
    This isolates the dependencies into a two-step procedure. Firstly, the PINN computes the outputs across a scan of $J_N^*$. The optimal configuration would be the one that matches the empirical ratio for each given value of $\lambda^*$:
    \begin{equation}
        \frac{M_N r_N}{J_N}\approx 3.957,
    \end{equation}
    using the values $M_N\approx\SI{939}{\MeV}$, and $r_N\approx\SI{0.72}{\fm}$ ($\hbar c\approx \SI{197.3}{\MeV\fm}$) \cite{ParticleDataGroup:2024cfk}.
    Then, having fixed $J_N^*$ for a given value of $\lambda^*$, the physical parameters are then obtained via algebraic substitution:
    \begin{equation}
        R_B=\frac{r_N}{r^*},\quad f=\frac{1}{R_B}\left(\frac{J_N}{J_N^*}\right)^{1/4},\quad \lambda=f^2 R_B^2\lambda^*.
     \end{equation}
    Now we can perform this methodology, fix the free parameters $\{f,R_B,J^*\}$ for different values of $\lambda^*$, and obtain, for instance, the prediction for the mass $M_\Delta$ of the Delta resonance ($j_\Delta=3/2$, $J_\Delta=\sqrt{15/4}$). The results are summarized in table \ref{tab:fit}.
    \begin{table}
        \centering
        \begin{tabular}{|c|c|c|c|c|c|}
            \hline
            $\lambda^*$ (input) & $J_N^*$ & $R_B$ ($\unit{\fm}$) & $f$ ($\unit{\MeV}$) & $\lambda$ & $M_{\Delta}$ ($\unit{\MeV}$) \\
            \hline
            0.01 & 0.049 & 77.5 & 5.2 & 0.04 & 938.0 \\
            0.10 & 0.918 & 5.6 & 34.9 & 0.10 & 949.4 \\
            0.50 & 1.920 & 5.7 & 28.2 & 0.33 & 998.5 \\
            0.80 & 5.720 & 1.8 & 67.1 & 0.31 & 1155.1 \\
            1.00 & 7.390 & 1.5 & 75.9 & 0.34 & 1032.1 \\
            \hline
        \end{tabular}
        \caption{Results of the phenomenological fit for the free parameters $f$, $R_B$ and $J^*$ given a previously fixed value of $\lambda^*$. The mass of the Delta resonance stands as a prediction.}
        \label{tab:fit}
    \end{table}
    From these results, we see that the limit $\lambda\to 0$ seems to be a poor approximation for describing the hadronic spectra since the prediction for $M_\Delta$ is far from the experimental value ($M_\Delta\approx\SI{1232}{\MeV}$ \cite{ParticleDataGroup:2024cfk}).  In fact, we obtain $M_\Delta\approx M_N$ in this limit. This is due to the fact that the mass splitting goes as $\propto J^2/\mathcal{I}$, so a huge moment of inertia would cause the mass splitting to collapse. As we will see, the limit $\lambda\to0$ precisely generates a centrifugal barrier that increases the moment of inertia. On the other hand, we have found that the fit that better predicts the mass of the Delta resonance corresponds to $\lambda^*\approx 0.8$. 
    
    However, we obtain $J_N^*=5.72$ for $\lambda^*=0.8$, so we have $J_\Delta^*=\sqrt{5}J_N^*=12.8$. These are values larger than unity, thus the convergence of the series \eqref{eq:energy} is not guaranteed. It seems from our results that the Delta resonance cannot be accurately described within our perturbative expansion.

\begin{figure}
        \centering
        \includegraphics[width=\linewidth]{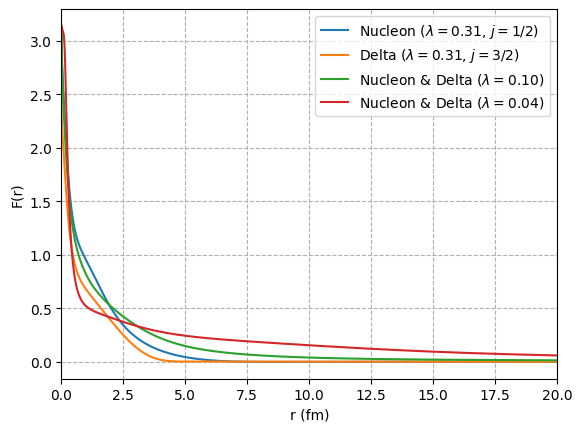}
        \caption{Brane-Skyrmion profiles of a Nucleon and Delta resonance for different values of $\lambda$ in physical units. The profiles for small $\lambda$ are mostly identical for different values of $j$.}
        \label{fig:fit}
    \end{figure}

    Finally, let us take a look at the profiles for different values of $\lambda$ and $j$ ($J=\sqrt{j(j+1)}$) in physical units. Since $F(r)$ is dimensionless, we only have to rescale the radial coordinate: $r=R_B r^*$. The results are shown in figure \ref{fig:fit}.
    We notice that for small $\lambda$ the profiles for different values of $j$ are mostly identical, denoting the fact that the mass splitting is constrained. When $\lambda$ is big enough, it is possible to distinguish both profiles, such as for $\lambda=0.34$ ($\lambda^*=1$).

    On the other hand, it is particularly noteworthy that for $\lambda\to 0$ the energy tends to be more concentrated near the origin. We can see this from the fact that the energy density goes as $F'(r)$ \eqref{eq:E_dens}, so there is more energy where $F(r)$ is steeper (see red curve in Figure \ref{fig:fit}). Essentially, this means that for $\lambda\to 0$ the dynamic solution tends to be highly localized near $r=0$, while for greater $\lambda$ the solutions tend to disperse more. In fact, we saw in \ref{Sec.II} that a static Brane-Skyrmion is point-like for $\lambda\to 0$. Now we have demonstrated that for $\lambda\to 0$ the dynamic solution tends to be highly localized, but not point-like, in contrast to the static case. This evasion of collapse can be understood via a scaling argument in the line of Derrick's theorem \cite{Derrick:1964ww}. Applying the scaling $r=L\tilde{r}$ to the Hamiltonian (\ref{eq:quantum-H}) and taking $L\to 0$ (in the $\lambda=0$ case), we have (\ref{eq:L2n}):
    \begin{align}
        L_{2n}(L)=\frac{4\pi}{4n^2-1}
        &\int_0^\infty\dd{\tilde{r}}L f^4\sqrt{1+\frac{R_B^2}{L^2}F'^2(\tilde{r})} \nonumber\\
        &\times\left(\frac{v^2}{f^4}\frac{L^2\tilde{r}^2\sin^2 F(\tilde{r})}{L^2\tilde{r}^2+R_B^2\sin^2 F(\tilde{r})}\right)^n.
    \end{align}
    For $n=0$, we recover the static $L\to 0$ limit: $H_0=2\pi^2 f^4 R_B^3$. However, for $n\geq 1$, Derrick's scaling dictates $L_{2n}\to 0$. Crucially, since $L_2=\beta\to 0$ represents the moment of inertia, the rotational kinetic energy ($\propto J^2/\beta$) diverges. This angular momentum generates a powerful centrifugal barrier, strictly preventing collapse and dynamically forcing the Brane-Skyrmion to adopt a finite, non-zero size and an energy greater than $2\pi^2 f^4R_B^3$.

    As already mentioned, the same argument explains the mass splitting for small values of $\lambda$. While $J$ defines specific quantum states, the induced curvature ($\lambda\neq0$) provides the dominant energy contribution. Although regimes with $J^*\gtrsim 1$ are possible, one must rigorously verify that successive terms in the series expansion decrease monotonically, a condition not universally guaranteed for arbitrary configurations, as we already have seen.
    
\section{Conclusions}
    This work establishes the canonical quantization of Brane-Skyrmions. We first demonstrated that the standard Atiyah–Manton ansatz is not generically a good approximation, even in purely static configurations, in contrast with the conventional Skyrme model (with massless pions). Also, by promoting the global symmetries to time-dependent collective coordinates, we have performed the rigid rotor quantization of these objects. Due to the highly nonlinear structure of the brane effective action, the Legendre transform required to obtain the Hamiltonian is technically involved. To overcome this, we have developed a perturbative scheme in which the transform is carried out order by order in an expansion in the total angular momentum operator. This approach allows us to systematically compute the quantum corrections to the energy arising from the quantized (iso)rotational modes of the classical soliton in the brane model, which is one of the main results of the paper

    In order to find solutions for the radial profile both in the classical and quantum regimes, we have implemented a Physics-Informed Neural Network (PINN) framework to autonomously minimize the complex energy functionals without the need to compute the equations of motion.
    
   Moreover, we support our numerical findings via scaling arguments, which physically explain how the induced centrifugal barrier prevents the collapse of the quantum solitons, and also how classical solitons can be point-like defects. Indeed, in the standard Skyrme model, the quartic term is essential for stability. Indeed, a simple scaling analysis implies that without it the soliton collapses to zero size \emph{and} energy. A similar argument applies in the brane setup with only the Nambu–Goto term, but with a crucial difference: the energy remains finite in the zero-size limit. In this regime, the system appears to become of BPS type \cite{Bogomolny:1975de}, since the $n_W=2$ hedgehog mass equals exactly twice the single skyrmion mass \cite{Cembranos:2001rp}. This allows for a parametric separation between total mass and binding energies, controlled by $\lambda$. This feature is not found in the standard Skyrme model, which typically predicts nuclear binding energies wrong by an order of magnitude.  
   Therefore, studying multi-skyrmion binding energies as a function of $\lambda$ is an interesting direction that we leave for future work.

    In the last section, we performed a phenomenological fit to model physical nucleons, as the original Skyrme model does. However, we have found that states with large effective spin parameter $J^*$ are not well suited for a description within our formalism, as the series expansion becomes non-convergent for these states. Developing a fully non-perturbative approach to the quantization could yield a more accurate mass spectrum.

    Despite these limitations, our Brane-Skyrmion model presents a significant theoretical counterpart to the standard Skyrme model. While the classical Skyrme approach only reaches the lowest order in $J^2$ for the rotational kinetic energy, the Dirac-Nambu-Goto nature of our effective action provides a systematic expansion to arbitrary higher orders in $J^2$. 
    Furthermore, while the standard Skyrme equations can be solved using traditional shooting methods, the analytical complexity of Brane-Skyrmions makes such approaches prohibitive. The implementation of PINNs in this work not only bypassed these analytical obstacles but also highlighted the immense potential of neural networks as unsupervised solvers in soliton models.

    Although there remains room for improvement, this work establishes a robust framework for exploring topological solitons within the context of extra-dimensional braneworld-inspired theories. Indeed, by changing the dimensionality of the extra dimensional space, or adding gauge fields \cite{Goon:2012mu}, a similar geometric effective action could admit other kinds of solitons such as vortices or monopoles.
    
    On the other hand, having successfully formulated the semiclassical quantization of a single Skyrmion, the model considered here opens numerous avenues for future research. These include introducing a potential for the (pseudo-)Goldstone fields, which is required for a realistic phenomenological description of pions and nuclei \cite{Manton:2022fcb}. In our framework, such a potential could naturally arise from a warp factor in the bulk metric, as demonstrated in \cite{Cembranos:2001rp}. Another interesting direction would be modeling the internal structure of dense nuclear matter as a periodic configuration \cite{Castillejo:1989hq,Adam:2022aes,Harland:2023ved} to compute the associated equation of state of neutron stars. This would mirror the approach recently developed within the (generalized) Skyrme model \cite{Adam:2020yfv,Adam:2021gbm,Adam:2022cbs,Adam:2023cee,Leask:2023tti}.

    \acknowledgments
    The authors would like to thank Christoph Adam, J.J. Blanco-Pillado and Bjarke Gudnasson for useful comments. The work of A.G.M.-C. is supported by Grants Nos. ED481B-2025/059 and ED431B-2024/42 (Xunta de Galicia postdoctoral fellowship). The work of J.A.R.C. is supported by project PID2022-139841NB-I00, funded by MICIU/AEI/10.13039/501100011033 and by ERDF/EU. The work of S.S.R. is supported by the ``Ayudas de Máster IPARCOS-UCM/2025'' grant. This work has also been carried out within the framework of the COST (European Cooperation in Science and Technology) Actions CA21106, CA21136, CA22113, and CA23130.


\bibliography{refs_draft} 

\appendix

\section{Shooting Method}\label{appendix:A}
    The shooting method is a numerical technique 
    designed to solve second-order boundary value problems where two conditions are specified for a function $F(r)$, but no initial condition is provided for its derivative $F'(r)$. 

    For the specific case of static, classical Brane-Skyrmions, we must solve the nonlinear differential equation that is obtained by extremizing the radial energy functional:
    \begin{equation}
        M_S[F]=-4\pi\int_0^\infty\dd{r}\mathcal{L}_\text{eff}[F,F',F'';r],
    \end{equation}
    where $\mathcal{L}_\text{eff}=(f^4-\lambda f^2 R)\sqrt{g}$. Note that the vacuum energy term $M_S[0]$ can be omitted, as it constitutes a constant shift and is thus irrelevant to the resulting equations of motion.

    At first glance, one could think that since the scalar curvature $R$ introduces second derivatives ($F''$) into the Lagrangian, one might expect the corresponding Euler-Lagrange equations to be of fourth order. However, the term $R\sqrt{g}$ corresponds precisely to the Einstein-Hilbert action for the induced metric. It is a well-established result that the second derivatives within this term can be integrated by parts, leaving only a boundary contribution. Consequently, the variational principle yields a strictly second-order, non-linear ordinary differential equation (ODE) for the profile $F(r)$. The explicit (second order) differential equation has been derived symbolically using  Mathematica, but its expression is too long to be illustrative, so we have omitted it. The corresponding \texttt{Mathematica} notebook has been made accesible at \cite{Github_code}.

    We implement the shooting method by defining the first derivative at the origin as the shooting parameter $s$: $F'(0)=-s$. Knowing the boundary condition $F(0)=\pi$, we approximate the solution near the origin as $F(r)\approx \pi-s r$ for sufficiently small $r>0$ (avoiding $r=0$ to prevent numerical singularities). This step initializes the integration. The ODE is then integrated outward, yielding a parametric solution dependent on $s$. Finally, the parameter $s$ is iteratively adjusted until the asymptotic boundary condition is satisfied: $F(\infty)=0$.
    
    This numerical procedure has been implemented in the Wolfram \texttt{Mathematica} notebook \cite{Github_code}. It has been used as an independent benchmark to validate the accuracy of the PINN approach, allowing for a direct comparison of the solutions, as illustrated in Figures \ref{fig:NN_1} and \ref{fig:NN_001}.

    \hide{
    \begin{figure}
        \centering
        \includegraphics[width=\linewidth]{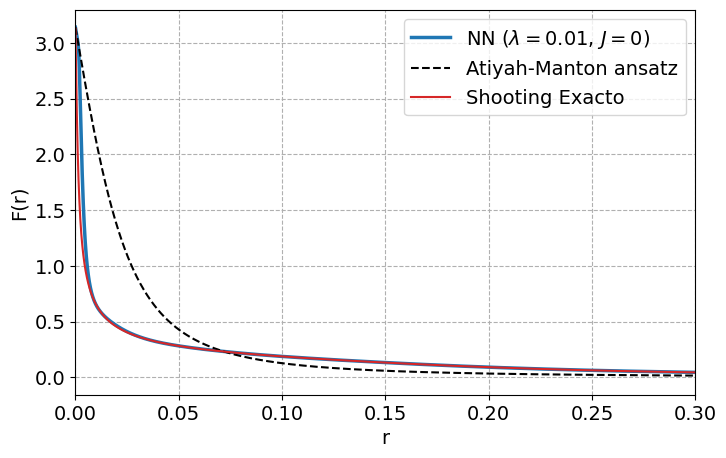}
        \caption{Comparison between the Atiyah-Manton solution, the shooting method solution, and the Neural Network solution for a static Brane-Skyrmion with $\lambda=0.01$. The shooting parameter is found to be $s\approx 1211$.}
        \label{fig:shooting_001}
    \end{figure}
    \begin{figure}
        \centering
        \includegraphics[width=\linewidth]{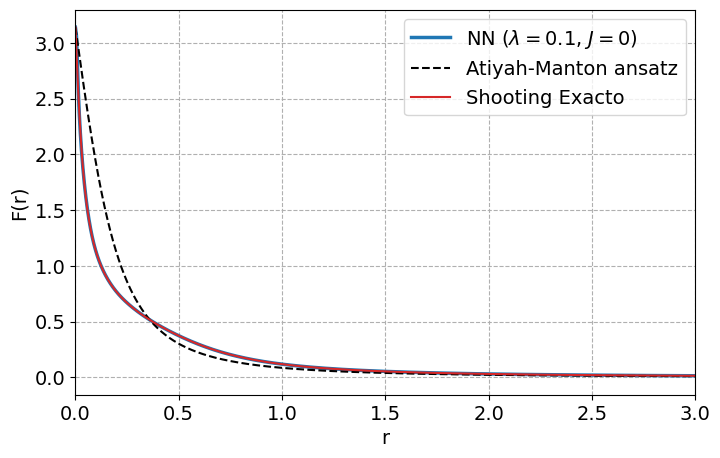}
        \caption{Comparison between the Atiyah-Manton solution, the shooting method solution, and the Neural Network solution for a static Brane-Skyrmion with $\lambda=0.1$. The shooting parameter is found to be $s\approx 44$.}
        \label{fig:shooting_01}
    \end{figure}
    }
    
    We see that both solutions provided by the shooting method and the PINN mostly coincide, which justifies the use of PINNs over traditional numerical methods since they are significantly more versatile, as they do not require explicit knowledge of the equations of motion. Indeed, the NN performs the minimization autonomously via automatic differentiation directly on the energy functional (see Appendix \ref{appendix:B}), bypassing the need to analytically derive the highly nontrivial Euler-Lagrange equations associated with the induced metric curvature. This capability proves invaluable during quantization, where the effective Hamiltonian lacks a straightforward density representation.

\section{Neural network (NN)}\label{appendix:B}
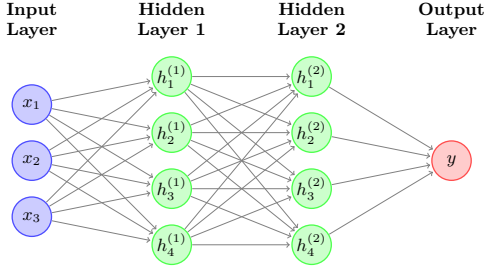
\begin{figure}
    \centering
    \resizebox{0.8\linewidth}{!}{%
    \begin{tikzpicture}[shorten >=1pt, ->, draw=black!50, node distance=2.5cm, scale=1.2, transform shape]
        
        \tikzstyle{every pin edge}=[<-,shorten <=1pt]
        \tikzstyle{neuron}=[circle, fill=black!25, minimum size=20pt, inner sep=0pt, font=\small]
        \tikzstyle{input neuron}=[neuron, fill=blue!20, draw=blue!70, thick]
        \tikzstyle{hidden neuron}=[neuron, fill=green!20, draw=green!70, thick]
        \tikzstyle{output neuron}=[neuron, fill=red!20, draw=red!70, thick]
        \tikzstyle{annot} = [text width=5em, text centered, font=\bfseries\small]

        \def\layersep{2.5cm}

        \foreach \name / \y in {1,...,3}
            \node[input neuron] (I-\name) at (0,-\y) {$x_\y$};

        \foreach \name / \y in {1,...,4}
            \path[yshift=0.5cm]
                node[hidden neuron] (H1-\name) at (\layersep,-\y) {$h^{(1)}_\y$};

        \foreach \name / \y in {1,...,4}
            \path[yshift=0.5cm]
                node[hidden neuron] (H2-\name) at (2*\layersep,-\y) {$h^{(2)}_\y$};

        \node[output neuron] (O) at (3*\layersep,-2) {$y$};

        \foreach \source in {1,...,3}
            \foreach \dest in {1,...,4}
                \path (I-\source) edge (H1-\dest);

        \foreach \source in {1,...,4}
            \foreach \dest in {1,...,4}
                \path (H1-\source) edge (H2-\dest);

        \foreach \source in {1,...,4}
            \path (H2-\source) edge (O);

        \node[annot, above of=H1-1, node distance=1cm] (hl1) {Hidden Layer 1};
        \node[annot, above of=H2-1, node distance=1cm] (hl2) {Hidden Layer 2};
        \node[annot, left of=hl1] {Input Layer};
        \node[annot, right of=hl2] {Output Layer};
        
    \end{tikzpicture}%
    }
    \caption{Generic multi-layer neural network diagram.}
    \label{fig:neural-network}
    \end{figure}

    Let us briefly outline the architecture of a generic NN, as illustrated in Figure \ref{fig:neural-network}. The general structure of the network consists of an \textbf{input layer}, or column vector representing the features of a single dataset (e.g., spatial coordinates). It contains the initial variables to be processed by the network.
    The dataset is then processed by a set of \textbf{hidden layers:} each layer applies a linear transformation to the output of the previous layer, followed by a non-linear activation function. Mathematically, the output of a layer is given by $\mathbf{h} = \sigma(\mathbf{W}\mathbf{x} + \mathbf{b})$, where $\mathbf{W}$ is the weight matrix, $\mathbf{b}$ is the bias vector, and $\sigma$ is the non-linear activation function. This non-linearity allows the network to model complex phenomena. 
 
 Finally, the \textbf{output layer} yields the final prediction of the model after all transformations through the hidden layers have been executed.

    Having established the architecture of the NN, we now outline the learning process:
    \begin{enumerate}
        \item \textbf{Forward pass:} The input data is fed into the NN, and its prediction is generated as described above.
        \item \textbf{Loss function:} In supervised learning, the prediction is compared against a known expected value using a \emph{loss function}, which quantifies the error. 
        \item \textbf{Backpropagation:} Using the chain rule, the gradient of the loss function with respect to each parameter of the network is computed.
        \item \textbf{Optimization:} An optimization algorithm updates the weights and biases based on the computed gradients to reduce the error (i.e., minimize the loss function) in the subsequent iteration. The iterative process terminates when the loss function converges to an acceptable minimum or a predefined tolerance criterion is met.
    \end{enumerate}

    Therefore, training a NN is fundamentally an optimization problem. In the context of PINNs, the input is a set of spatial coordinates $\mathbf{x}$, and the output is a continuous function $f(\mathbf{x})$. Instead of relying on labeled data, the traditional loss function is replaced by a physical functional $\mathcal{F}[f]$, such as the energy of the system. Thus, the network learns to find the physical configuration that minimizes the energy without requiring prior experimental data.

\section{Hamiltonian coefficients}\label{appendix:C}
    Recall that the classical Hamiltonian for spherical Brane-Skyrmions can be expressed as a power series in $J^2$:
    \begin{equation}
        H(J)=\sum_{n=0}^\infty H_{2n}J^{2n}=\alpha+\frac{1}{4\beta}J^2-\frac{\gamma}{16\beta^4}J^4+\mathcal{O}(J^6),
    \end{equation}
    provided that $H_{2n} J^{2n}\gg H_{2(n+1)}J^{2(n+1)}$, which ensures the convergence of the series in the slow-rotation regime.

    The parameters $\alpha$, $\beta$, $\gamma$, and subsequent terms represent the coefficients of the classical Lagrangian expansion:
    \begin{equation}
        L(\omega)=\sum_{n=0}^\infty L_{2n}\omega^{2n}=-\alpha+\beta \omega^2+\gamma\omega^4+\mathcal{O}(\omega^6).
    \end{equation}
    These can be derived by substituting the expansions of (\ref{eq:dynamic-metric}) and $R$:
    \begin{equation}
        \sqrt{g}=\sqrt{g_0}\sum_{n=0}^\infty A_n C^n(r)\sin^{2n}\theta\omega^{2n},\quad R=\sum_{n=0}^\infty R_{2n}\omega^{2n},
    \end{equation}
    where
    \begin{equation}
        A_0=1,\quad A_n=-\frac{(2n-2)!}{2^{2n-1}n!(n-1)!},\quad n\geq 1,
    \end{equation}
    and where $R_{2n}$ denote the coefficients of the scalar curvature expanded in powers of $\omega^2$, which can be computed symbolically using Wolfram \texttt{Mathematica} \cite{Github_code}.

    Note that $A_0$ is strictly positive, whereas $A_n$ is negative for $n\geq 1$; this sign alternation explains the definitions $L_0=-\alpha$ and $L_{2n}=\{\beta,\gamma,\dots\}$ for $n\ge 1$. By combining these expansions with the total Lagrangian (\ref{eq:dynamic-lagrangian}) and grouping terms by order, we obtain the following general expression:
    \begin{align}
        L_{2n}=\int\dd[3]{x}&\sqrt{g_0}
        \Big[A_n(-f^4+\lambda f^2 R_0)C^n(r)\sin^{2n}\theta \nonumber\\
        &+\lambda f^2\sum_{k=0}^{n-1}A_k R_{2n-2k}C^k(r)\sin^{2k}\theta\Big],
    \end{align}
    where $r\in[0,\infty)$, $\theta\in[0,\pi)$, and $\phi\in[0,2\pi)$; additionally, $\sqrt{g_0}$ is given by (\ref{eq:static-det}) and $C(r)$ by (\ref{eq:C(r)}).
\hide{
    It is particularly insightful to evaluate these coefficients in the $\lambda=0$ limit:
    \begin{align}
        \alpha(\lambda=0)&=4\pi\int_0^\infty\dd{r}f^4\frac{\sqrt{g_0}}{\sin\theta},
\\
        \beta(\lambda=0)&=\frac{4\pi}{3}\int_0^\infty\dd{r}f^4\frac{\sqrt{g_0}}{\sin\theta}C(r),
\\
        \gamma(\lambda=0)&=\frac{4\pi}{15}\int_0^\infty\dd{r}f^4\frac{\sqrt{g_0}}{\sin\theta}C^2(r),
    \end{align}
    }
    It is particularly insightful to evaluate these coefficients in the $\lambda=0$ limit, in which they admit the following closed-form expression:
    \begin{equation}\label{eq:L2n}
        L_{2n}=\frac{4\pi}{4n^2-1}\int_0^\infty\dd{r}f^4\frac{\sqrt{g_0}}{\sin\theta}C^n(r).
    \end{equation}
     where we note that the quotient $\sqrt{g_0}/\sin\theta$ is independent of $\theta$ and solely a function of $r$.
    To rigorously analyze the convergence of the effective Lagrangian series, one must examine the dimensionless expansion parameter, $\omega^2 C(r)$ (by definition (\ref{eq:C(r)}), $C(r)$ has dimensions of length squared). Since we assume $\omega\ll 1$, and given that $C(r)$ is bounded over the radial domain ($C(r)\to 0$ as $r\to 0$ and $r\to\infty$), the condition $\omega^2 C(r)\ll 1$ holds uniformly. Consequently, consecutive terms in the Lagrangian density scale as $(\omega^2 C)^{n+1}\ll (\omega^2 C)^n$, leading to $L_{2(n+1)}\omega^{2(n+1)}\ll L_{2n}\omega^{2n}$. Therefore, the power series expansion is robustly convergent for the physical scenarios of interest, fully justifying its truncation at the first few orders in $J^2$.

\end{document}